\documentclass[a4paper,fleqn, preprint, sort&compress]{cas-dc}

\usepackage{graphicx}
\usepackage{dcolumn}
\usepackage{bm}
\usepackage{hyperref}
\usepackage{slashed}
\usepackage[dvipsnames]{xcolor}
\usepackage{float}
\usepackage{multirow}
\usepackage{amsmath}
\usepackage{amsfonts}

\usepackage[numbers]{natbib}

\usepackage[final]{changes}
\definechangesauthor[color=orange]{Jan}
\definechangesauthor[color=green!50!black]{Wim}
\definechangesauthor[color=blue]{rev}

\newcommand{\carbon}{^{12}\text{C}}
\newcommand{\calcium}{^{40}\text{Ca}}
\newcommand{\calciumN}{^{48}\text{Ca}}

\def\threej#1{\inthreej(#1)}
\def\inthreej(#1,#2,#3,#4,#5,#6){\setlength\arraycolsep{2pt}\begin{pmatrix}#1 & #2 & #3 \\ #4 & #5 & #6 \end{pmatrix}}

\def\tsc#1{\csdef{#1}{\textsc{\lowercase{#1}}\xspace}}
\tsc{WGM}
\tsc{QE}

\begin{document}
\let\WriteBookmarks\relax
\def\floatpagepagefraction{1}
\def\textpagefraction{.001}

\shorttitle{Phase-space distributions of nuclear short-range correlations}    

\shortauthors{W. Cosyn, J. Ryckebusch}  

\title [mode = title]{Phase-space distributions of nuclear short-range correlations}  

\author[1,2]{W.~Cosyn}[orcid=0000-0002-9312-8569]
\cormark[1]
\ead{wcosyn@fiu.edu}

\author[2]{ J.~Ryckebusch}[orcid=0000-0001-7750-1522]
\ead{Jan.Ryckebusch@ugent.be}

\cortext[1]{Corresponding author}
%

\affiliation[1]{organization={Department of Physics, Florida International University},
            city={Miami},
          citysep={}, 
            postcode={33199}, 
            state={FL},
            country={USA}}

\affiliation[2]{organization={Department of Physics and Astronomy, Ghent University},
            city={Gent},
            postcode={B9000}, 
            country={Belgium}}

\begin{abstract}
Nuclear short-range correlations (SRCs) induce high-momentum/high-energy fluctuations in the nuclear medium.  In order to assess their impact on nuclear bulk properties, like nuclear radii and kinetic energies, it is instrumental to determine how SRCs are distributed in phase space as this sheds light on the connection between their appearance in coordinate and momentum space.  Using the lowest-order correlation operator approximation (LCA) to include SRC, we compute two-dimensional nuclear Wigner quasiprobability distributions $w(r, k)$ to locate those $({r}, {k})$ phase-space regions that are most heavily impacted by SRCs. The SRC-induced high-momentum components find their origin in a radial range that is confined to the nuclear interior. Significant SRCs strength is generated in the full momentum range $0 \leq k \lesssim 5 ~\text{fm}^{-1} $ covered in this work, but below the Fermi momentum those are dwarfed by the mean-field contributions. As an application of $w(r, k)$, we focus on the radial dependence of the kinetic energy $T$ and the momentum dependence of the radius $r_{\text{rms}}$ for the symmetric nuclei $\carbon$, $\calcium$ and the asymmetric nucleus $\calciumN$.  
The kinetic energy almost doubles after including SRCs, with the largest increase occurring in the nuclear interior $r \lesssim 2$ fm.  The momentum dependence of the $r_{\text{rms}}$ teaches that the largest contributions stem from  $k \lesssim 2 $ fm$^{-1}$, where the SRCs induce a slight reduction of the order of a few percent. The SRCs systematically reduce the  $\calciumN$ neutron skin by an amount that can be 10\%.
\end{abstract}


\maketitle

\section{Introduction}
\label{sec:introduction}


The size of an atomic nucleus \cite{Hagen:2015yea,Ruiz:2016gne} and how protons and neutrons are spatially arranged for various proton-to-neutron ratios \cite{Reed:2021nqk,Adhikari:2021phr,Atkinson:2019bwd,CREX:2013} are topics of continued great interest in the precision era of nuclear physics. Detailed nuclear-structure studies have shown that long-range correlations connected with core-breaking effects  have a substantial impact on computed proton and neutron nuclear radii and are sources of uncertainties in advanced nuclear-structure calculations.  For example, a systematic study for $\calciumN$ \cite{Hagen:2015yea} indicated that in \textit{ab initio} theory with a   family of modern chiral effective forces, the variations in the computed proton and neutron radii can be of the order of 10\%.  The impact of short-range correlations (SRCs) on bulk nuclear properties like radii is not that well known and has recently been addressed in \cite{Miller:2018mfb}.  In that paper, qualitative arguments are developed as to why the omission of SRCs may have a non-negligible impact on the computed radii of neutron-rich nuclei and it was suggested that more quantitative calculations are in order.

Wigner distributions~\cite{Wigner:1932eb, Hillery:1983ms} provide a distinct view on the spatial and momentum structure of quantum systems and are widely applicable including in subatomic physics~\cite{Weinbub:2018}. It is a subject of great interest in non-perturbative quantum chromodynamics (QCD)~\cite{Ji:2003ak, Belitsky:2003nz,Meissner:2008ay,Meissner:2009ww,Lorce:2011kd,Lorce:2013pza,Bhattacharya:2017bvs,Hagiwara:2017fye}.  The QCD Wigner distribution provides information about the joint position-momentum distributions of partons in the nucleon and as such can be considered as the mother partonic distribution from which all others
can be derived. 
Selected results of Wigner distributions for finite nuclei have been presented in Refs.~\cite{Shlomo:1981ayz,Prakash:1981ilg,Bonasera:1993zz,Shlomo:2021hrd}. A recent calculation using realistic potentials highlighted the influence of SRCs on the deuteron's Wigner distribution~\cite{Neff:2016ajx}.

In this work we aim at providing a study of the position-momentum structure of SRCs by presenting calculations of Wigner distributions for finite nuclei. The SRCs have been connected with fat momentum tails in the nuclear momentum distributions~\cite{Feldmeier:2011qy, Atti:2015eda, Hen:2016kwk, Lonardoni:2017egu}. The spatial structure of the SRCs in finite nuclei has received less attention and will be one of the topics of discussion here.
To quantify the impact of SRCs, we use the lowest-order correlation operator approximation (LCA)~\cite{Vanhalst:2014cqa, Ryckebusch:2018rct, Ryckebusch:2019oya} that is based on a number of assumptions: (i) the scale separation between the long-range and short-range nuclear correlations; (ii) the universal local character that make SRCs a property that can be imposed on the mean-field behavior through the operation of universal operators~\cite{Bogner:2012zm,Rios:2013zqa,Tropiano:2021qgf}. The LCA shares these assumptions with alternate theoretical approaches to quantify the impact of SRCs, including the generalized contact formalism (GCF) \cite{Weiss:2015mba,Cruz-Torres:2019fum}.  As LCA shifts the complexity from the wave functions to the operators, it can be used for SRCs estimation in nuclear structure and nuclear reaction applications. Since the LCA formalism does not account for long-range correlations, our focus is on the \textit{relative} contribution of SRCs to nuclear bulk properties like point-nucleon radii and kinetic energies. We deem that our calculations serve as an important comparative benchmark for highlighting the impact of SRCs on nuclear bulk properties.

The kinetic energy is connected with the momentum structure whereas the radii are connected with the spatial structure. With the Wigner distributions one gains access to the momentum structure of radii and the spatial structure of the kinetic energy, and how those are impacted by SRCs. Alternate approaches \cite{Cruz-Torres:2019fum} have addressed the SRCs in both coordinate and momentum space. Wigner distributions provide a unique window of insight into the phase-space distributions as they  provide information in both variables simultaneously.

In what follows, we first develop a formalism to compute Wigner distributions that include the effect of SRCs. We then proceed with the presentation of the results of numerical calculations for the separated proton and neutron Wigner distribution for the nuclei $\carbon$, $\calcium$ and $\calciumN$.  Those distributions form the basis to elucidate the phase-space dependence of SRCs in the proton and neutron radii and kinetic energies.

\section{Formalism}
\label{sec:formalism}

The Wigner distribution is the central quantity of interest in this work.
The quantum Wigner operator has the following spectral decomposition in three-dimensional coordinate ($\equiv \bm r$) or momentum ($\equiv \bm k$) space: 
\begin{multline}
    \widehat{w}(\bm r,\bm k) = \frac{1}{(2\pi)^3}\int d\bm x \;e^{i\bm k \cdot \bm x} \; \Bigl| \bm r-\frac{\bm x}{2}\Bigr\rangle \Bigl\langle \bm r+\frac{\bm x}{2}\Bigr| \\ = \frac{1}{(2\pi)^3}\int d\bm q\; e^{-i\bm q\cdot \bm r} \;\Bigl| \bm k+\frac{\bm q}{2}\Bigr\rangle \Bigl\langle \bm k-\frac{\bm q}{2}\Bigr|\,.
\end{multline}
In the coordinate space representation, $\bm k$ is Fourier conjugate to the relative coordinate $\bm x$ of the off-diagonal spatial density, while $\bm r$ is Fourier conjugate to the relative momentum $\bm q$ in the off-diagonal momentum density.
The Wigner quasiprobability distribution  for a pure state $\Psi$ is the expectation value of the $\widehat{w}(\bm r,\bm k)$:
\begin{equation}
    w(\bm r,\bm k) = \langle \Psi | \,    \widehat{w}(\bm r,\bm k) \,| \Psi \rangle.
\label{eq:wofvecrandveck}
\end{equation}

The integral of the Wigner operator over $\bm r$ $(\bm k)$ yields the momentum (spatial) density operator
\begin{align}\label{eq:wigner_obmd}
    &\hat{n}(\bm k) = \int d\bm r \;\widehat{w}(\bm r,\bm k) = |\bm k\rangle \langle \bm k|\,,\\ 
    &\hat{\rho}(\bm r) = \int d\bm k \; \widehat{w}(\bm r,\bm k) = |\bm r\rangle \langle \bm r|\,.\label{eq:wigner_obrd} 
\end{align}
Expectation values of operators $\widehat{F}$ can be written as the phase-space average of the Wigner distribution 
\begin{equation} \label{eq:wigner_me}
    \langle \widehat{F} \rangle = \iint d\bm r d\bm k \; w(\bm r,\bm k)f(\bm r,\bm k)\,,
\end{equation}
where the Wigner transform $f(\bm r,\bm k)$ of $\widehat{F}$ is defined as
\begin{align}
    f(\bm r,\bm k) &\equiv \int d\bm x \; e^{-i\bm k\cdot \bm x} \; \Bigl\langle \bm r-\frac{\bm x}{2}\Bigr| \,\widehat{F} \,\Bigl| \bm r+\frac{\bm x}{2} \Bigr\rangle \, ,\nonumber\\
    &=\int d\bm q \; e^{i\bm q\cdot \bm r} \; \Bigl\langle \bm k+\frac{\bm q}{2}\Bigr| \,\widehat{F} \,\Bigl| \bm k-\frac{\bm q}{2} \Bigr\rangle\,.
\end{align}

In this work, we assume spherical symmetry and present results for the $w(r,k)$ which are obtained after integrating $w(\bm r,\bm k)$ of Eq.~(\ref{eq:wofvecrandveck}) over the solid angles of the spatial and momentum coordinates
\begin{equation}
    w(r,k) \equiv \iint d\Omega_{\bm r}\, d\Omega_{\bm k} \; w(\bm r ,\bm k )\, .
    \label{eq:wignerinrandk}
\end{equation}
The Wigner distribution $w(r,k)$ is the quasiprobability distribution in the nucleon's radial coordinate and momentum. It is normalized to the number of nucleons $A$ 
\begin{equation}
    \int r^2 dr \int k^2 dk \; w(r,k)  = A \, .
\end{equation}
An outline of the derivation of the Wigner distribution in the LCA is given in App.~\ref{app:lca_wigner}. Through the introduction of SRCs operators, many-body operators between Slater-determinant states are generated. In LCA, the many-body operators generated from one-body operators are truncated at the level of two-body operators  and the $w(r,k)$ can be separated in a proton and a neutron part by considering the four isospin pair combinations (see Eq.~(\ref{eq:w_LCA_operator}))
\begin{equation} \label{eq:wrkisospindecompostion}
    w(r,k) =  \left[ w_{pp}(r,k) + w_{pn}(r,k) \right] + \left[ w_{nn}(r,k) +  w_{np}(r,k) \right] \, .
\end{equation}
Hereby, the $w_{pn}$ has two categories of contributions. The first category stems from an uncorrelated proton and neutron that are both described as quasi-particles in the mean field. The second category is the SRCs one whereby the tagged proton and neutron are correlated through one or a product of two correlation operators. Similar discussions hold for the other three pair combinations.

To obtain the nuclear rms radius and kinetic energy, we consider scalar operators $\hat{r}^2,\hat{k}^2$ which have Wigner transforms $r^2,k^2$ respectively.  We can use Eq.~(\ref{eq:wigner_me}) to extract the radial dependence of the kinetic energy operator in coordinate space, and that of the rms radius in momentum space.  The first method permits to calculate the quasi-expectation value of $\hat{T}=\hat{k}^2/2m$ ($\hat{r}^2$) at a given position (momentum)
\begin{align}
    &T(r)=\Bigl< \widehat{T}(r) \Bigr> = \frac{\int k^2 dk\; \tfrac{k^2}{2m} \; w(r,k)  }{\int k^2dk\; w(r,k) }\,,\label{eq:exp_T}\\
    &r_\text{rms}(k) \equiv \sqrt{\langle \hat{r}^2(k) \rangle} = \sqrt{\frac{\int r^2 dr\; r^2 w(r,k)  }{\int r^2dr\; w(r,k) }}\label{eq:exp_r}\,.
\end{align}
Due to quantum effects in the Wigner distribution, these quasi-expectation values of positive definite operators can be negative.
These variables allow to infer the magnitude of the nucleon kinetic energy at a certain $r$, or the size of the nuclear rms radius  with a certain $k$.  Because of the $r$-($k$-) dependence in both numerator and denominator, the $T(r)$ ($r^2_\text{rms}(k)$) do not integrate to the full $T$ ($r^2_\text{rms}$). 

A distribution that is both reminiscent of the spatial structure of the kinetic energy (momentum structure of the nuclear radius) and provides the proper scalar quantity after integration, can be obtained in the second method that is based on the computation of the densities $\rho_T(r)$ and $\rho_{r^2}(k)$:
\begin{align}
    &\rho_T(r) \equiv \frac{r^2\int \,k^2 dk\; \tfrac{k^2}{2m} \; w(r,k)  }{\int r^2dr \int k^2dk\; w(r,k) }\,, \nonumber\\
     &\int dr \rho_T(r) = \bigl< \widehat{T} \bigr>=T\,;\label{eq:tkin_wigner}\\
    &\rho_{r^2}(k) \equiv \frac{k^2 \int r^2 dr\; r^2 w(r,k)  }{\int k^2dk\int r^2dr\; w(r,k) }\,,\nonumber\\
    &\int dk \rho_{r^2}(k) = \langle \hat{r}^2 \rangle = r_\text{rms}^2 \label{eq:rms_wigner}\,.
\end{align}
The $\rho_T(r)$ encodes the contribution to the kinetic energy at given nucleon radial coordinate $r$ and was also considered in Ref.~\cite{Prakash:1981ilg}. Analogously, $\rho_{r^2}(k)$ encodes the contribution to the nuclear radius squared at given momentum $k$. In this work we assess the impact of SRCs on the $T(r)\,(r_\text{rms}(k))$ and $\rho_T(r)\,(\rho_{r^2}(k))$, each offering complementary insight in what actually happens to bulk nuclear properties after including high-momentum/high-energy fluctuations in a  model.

%
\section{Results}
\label{sec:results}

The LCA has the following inputs: (i) a set of universal strength correlation functions $f_p (r_{12})$ entering the matrix elements of Eq.~(\ref{eq:LCA_me}); (ii) the HO frequency for which we adopt a global parameterization of the form
\begin{equation} \label{eq:HO_param}
    \hbar\omega = c_1 A^{-\frac{1}{3}} - c_2 A^{-\frac{2}{3}}\,.
\end{equation}
As in previous publications~\cite{Vanhalst:2014cqa,Ryckebusch:2018rct,Ryckebusch:2019oya}, we use the Argonne VMC correlation functions~\cite{Pieper:1992gr} for the tensor and spin-isospin correlation functions, and use two options for the central correlation function $f_c (r_{12})$: a hard one that is computed with the aid of the Reid potential (denoted $f_c[R]$)~\cite{Dickhoff:2004xx}, and the softer Argonne correlation function ($f_c[V]$)~\cite{Pieper:1992gr}.  We stress that with the $f_c[R]$ we obtain momentum distributions that are very similar to those obtained in \textit{ab initio} calculations \cite{Ryckebusch:2018rct}. Furthermore, the choice for $f_c[R]$ was a data driven one, as $^{12}$C$(e,e'pp)$ data could be described with this choice \cite{Blomqvist:1998gq}. For the HO frequency of Eq.~(\ref{eq:HO_param}), we have in previous LCA works systematically used the ``default'' values $c_1=45~\text{MeV}, c_2=25~\text{MeV}$, a choice that we refer to as $\hbar\omega[d]$.  As we quantify the effect of SRCs on nuclear radii in this work and wish to quantify uncertainties stemming from the model parameters, we also explore other values of $c_1,c_2$ by fitting them to the measured nuclear rms charge radii of $^{4}\text{He}$, $^{9}\text{Be}$, $^{12}\text{C}$, $^{16}\text{O}$, $^{27}\text{Al}$, $^{40}\text{Ca}$, $^{48}\text{Ca}$, $^{56}\text{Fe}$, $^{108}\text{Ag}$, $^{197}\text{Au}$ and $^{208}\text{Pb}$~\cite{Angeli:2013epw}.  This was done using a standard minimum $\chi^2$ fit.  These inputs carry the label $\hbar\omega[f]$. In this work, the IPM corresponds with the HO model that can be formally reached after setting all correlation operators equal to zero in LCA. We consider this HO model as the benchmark against which to measure the impact of SRCs.

\begin{figure}[htb]
    \centering
    \includegraphics[width=0.5\textwidth]{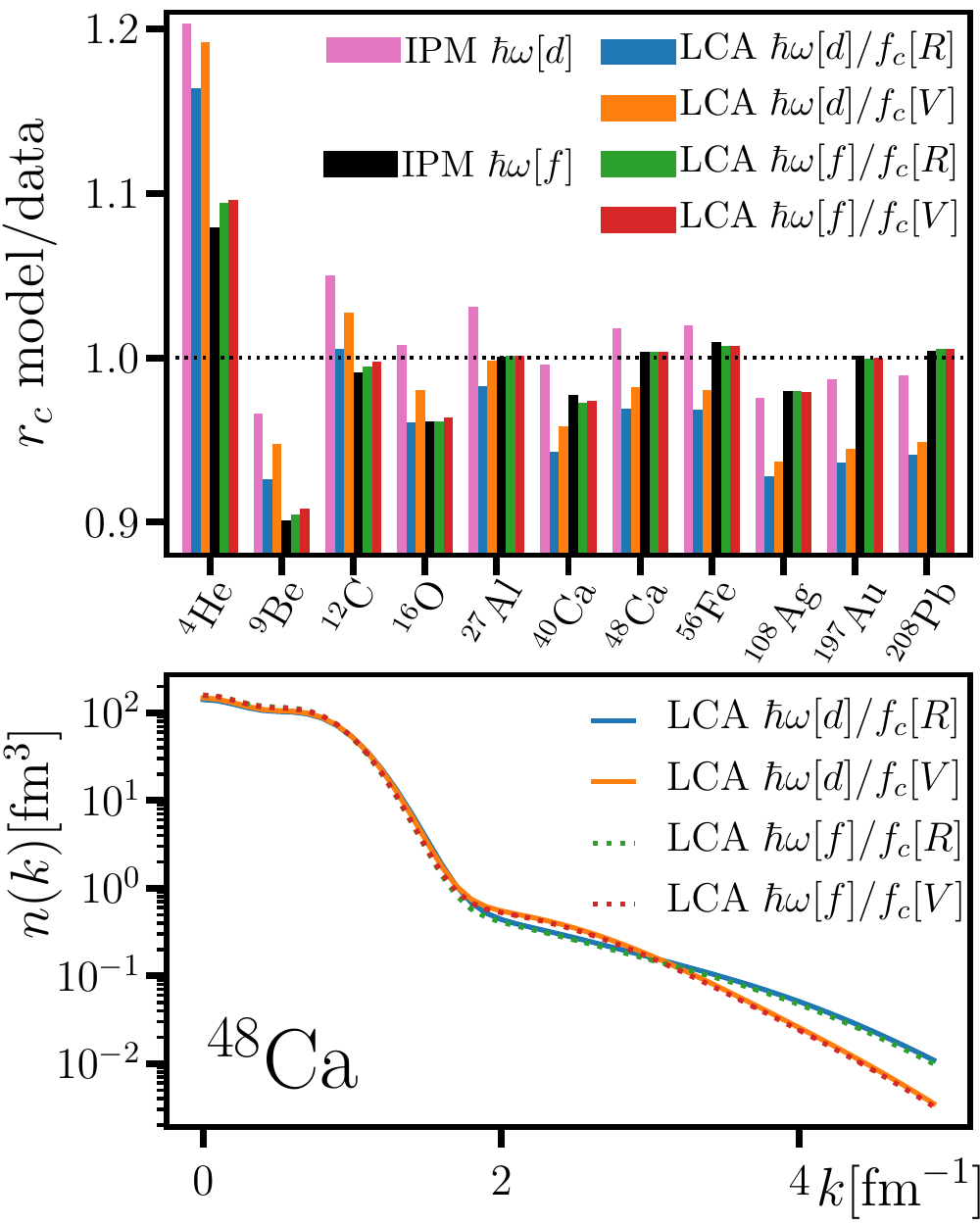}
\caption{Top: Ratio of model charge radii $r_c$ (including the point rms radius correction of ~\cite{Hagen:2015yea}) and data~\cite{Angeli:2013epw}. Bottom: One-body momentum distribution $n(k)$ for $^{48}$Ca as computed in LCA with the four model variants detailed in Table~\ref{tab:fits}. 
}
    \label{fig:rms_fit}
\end{figure}

\begin{table}[htb]
    \centering
    \caption{Summary of the model inputs used in this work.  Left column labels are explained in the text.}
    \label{tab:fits}
    \begin{tabular}{l|c|cc}
    \toprule
         Label & $f_c$ & $c_1$ [MeV] & $c_2$ [MeV]  \\
    \midrule     
        IPM $\hbar\omega[d]$ & none & 45 & 25\\ 
        IPM $\hbar\omega[f]$ & none & $40.0 \pm 1.2$ & $2.8 \pm 5.1$\\ 
        LCA $\hbar\omega[d]/f_c[R]$ & Reid & 45 & 25\\
        LCA $\hbar\omega[f]/f_c[R]$ & Reid & $36.7  \pm 1.1$ & $4.1 \pm 4.8$\\
        LCA $\hbar\omega[d]/f_c[V]$ & VMC & 45 & 25\\
        LCA $\hbar\omega[f]/f_c[V]$ & VMC & $36.5 \pm 0.4$ & $0.01 \pm 1.0$\\
        \bottomrule
    \end{tabular}
\end{table}

As a robustness check and sensitivity analysis, we consider two choices for the two inputs to LCA. Together with the two HO frequency variants of the IPM we are left with six models that are listed in Table~\ref{tab:fits}. In Fig.~\ref{fig:rms_fit} we compare the computed charge radii to data for these models.
  As we compute point-nucleon radii, corrections (see Ref.~\cite{Hagen:2015yea}) were applied before comparing to data.  Note that the fits produce $c_1$ that are somewhat smaller than the default value and a small $c_2$. We stress that the variation of $\hbar \omega$ across the various model variants does not exceed 13\%. In Fig.~\ref{fig:rms_fit}, we see that the IPM $\hbar\omega[d]$ model provides reasonable agreement with the measured rms radii.  The LCA calculations with those HO parameters produce rms radii that are a few percent smaller whereby the strongest impact of the SRCs is observed for the LCA model that uses a hard central correlation function.   For medium and heavy nuclei, the fitted IPM and LCA $\hbar\omega[f]$ parameterizations  yield almost identical charge radii, in overall good agreement with the data. For the light nuclei there are larger deviations between the predicted radii obtained with the $\hbar\omega[d]$ and $\hbar\omega[f]$ parameters. With the $\hbar\omega[f]$ HO frequency the measured radius for $\carbon$ can be reproduced.  The larger deviations between computed and measured radii for $^4$He and $^9$Be can be attributed to the absence of long-range effects in the LCA. For these light nuclei, also the center-of-mass (c.o.m.)~corrections can be sizable \cite{Hagen:2009a}.

%
The numerical cost of LCA scales polynomially with mass number $A$ which makes it applicable throughout the mass table. In this work, our focus is on $\carbon$,   $\calcium$, and $\calciumN$. The latter nucleus allows us to assess the SRC effects in asymmetric nuclei. For $\calciumN$, the bottom panel in Fig.~\ref{fig:rms_fit} shows the LCA one-body momentum distribution for the four model variants.  
Contrary to the radii, the choice of HO frequency has an almost negligible effect on the momentum distribution. The choice for the central correlation function, on the other hand, mainly affects the momentum distribution for momenta larger than half the nucleon mass $k \gtrsim 3~\text{fm}^{-1}$.  As reported in Refs.~\cite{Ryckebusch:2018rct,Ryckebusch:2019oya}, as relatively little strength is present at those momenta this difference results in a variation of the order of a few percent in comparisons with quantities extracted from electron scattering data. For the LCA momentum distributions for $\carbon$ and $\calcium$ (shown in Ref.~\cite{Ryckebusch:2018rct}), similar remarks hold for the sensitivity to the two input sources as for $\calciumN$.

Over the last decade, constraints of the SRCs models have been improved through the increased availability of two-nucleon knockout data from  proton- and electron-nucleus experiments in selected kinematics~\cite{Tang:2002ww,Piasetzky:2006ai,Shneor:2007tu,Subedi:2008zz,Hen:2014nza,Korover:2014dma,Duer:2018sxh}.  Thanks to its flexibility, the LCA model could be well tested against results for the isospin \cite{Ryckebusch:2018rct} and mass \cite{Colle:2015ena, Ryckebusch:2019oya} dependence of SRCs. The LCA provides a good basis for accurate SRCs modeling across the nuclear mass table. As an illustration of this we mention the so-called $a_2$ scaling factors that are extracted from inclusive electron-nucleus data and can be connected to the aggregated effect of SRCs in nucleus $A$ relative to the deuteron. A recent $^{48}$Ca$(e,e^{\prime})$ / $^{40}$Ca$(e,e^{\prime})$ measurement has addressed the isospin structure of SRCs \cite{Nguyen:2020mgo} and provided the result $0.971 \pm 0.012$ for the $a_2(\calciumN)/a_2(\calcium)$ ratio of measured cross sections per nucleon. Loosely speaking, this implies that per nucleon there is about 3\% less impact from SRCs in  $^{48}$Ca as compared to $^{40}$Ca. The LCA predictions for this quantity (that were published before the data \cite{Ryckebusch:2019oya}) can be extracted from the high-momentum tails of the LCA momentum distribution for $\calcium$ and $\calciumN$. The computed numbers for the $^{48}$Ca / $^{40}$Ca   ratio are 4.89/4.99=0.98 (see Table~I of Ref.~\cite{Ryckebusch:2019oya}) which is in close agreement with the data.

\begin{figure*}[htb]
    \centering
    \includegraphics[width=\textwidth]{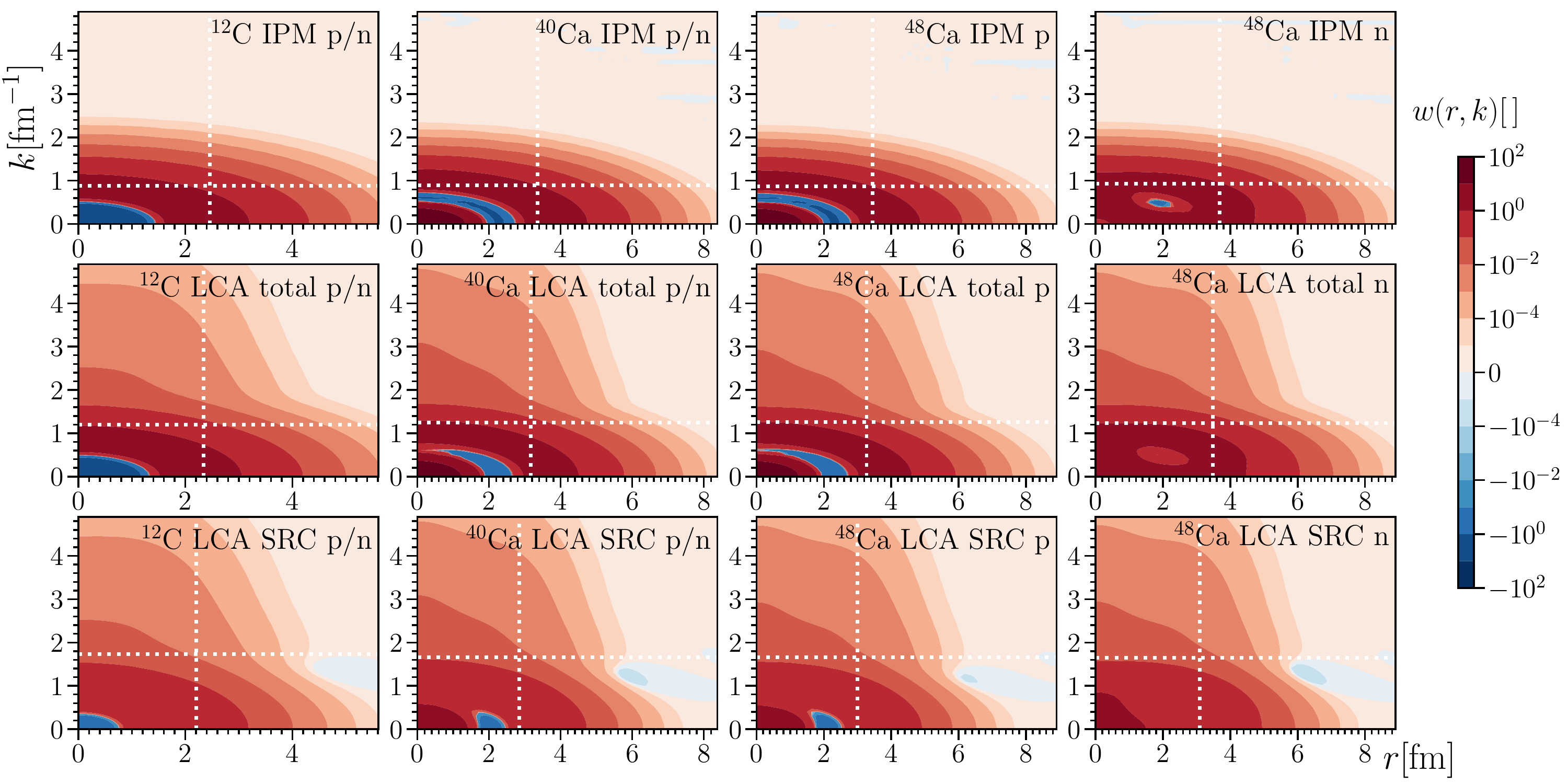}
    \caption{Two-dimensional proton and neutron Wigner distributions $w(r,k)$ for $\carbon$, $\calcium$, and $\calciumN$ as computed in the IPM and the LCA using the $\hbar\omega[d]/f_c[R]$ input. For the two symmetric nuclei, the proton (p) and neutron (n) result are identical.  Mean-field results (top row); full LCA results (middle row); SRCs contribution to LCA results (bottom row). For each of the 12 distributions, the white vertical  (horizontal) line denotes the corresponding $r_\text{rms}$ ($k_\text{rms}$). For the IPM and full LCA results (top and middle rows)  the $w(r,k)$ is normalized to the total number of protons/neutrons (6/6 for $\carbon$; 20/20 for $\calcium$; 20/28 for $\calciumN$), while the normalization of the bottom panel is equal to the LCA prediction for the number of SRCs protons/neutrons (1.8/1.8 for $\carbon$; 7.6/7.6 for $\calcium$; 8.0/9.9 for $\calciumN$). }
    \label{fig:wigner}
\end{figure*}

Figure~\ref{fig:wigner} shows  the results for the two-dimensional Wigner distribution $w(r,k)$.  
These LCA numerical results were subjected to several checks.  One-body momentum distributions $n(k)$ obtained using Eq.~(\ref{eq:wigner_obmd}) were confronted with those from direct computation~\cite{Ryckebusch:2019oya}.  Similarly, the rms radii  obtained with Eq.~(\ref{eq:rms_wigner}) that involves the $w(r,k)$ were compared with those obtained through direct calculations with the rms radius operator.  Finally, the normalization of the four isospin pair combinations $(pp,nn,np,pn)$ of the LCA  $w(r,k)$ of Eq.~(\ref{eq:wrkisospindecompostion}) are constrained by $Z$ and $N$.\footnote{ Note that for these normalizations this is not a trivial result, as we expand the denominator of the operator matrix elements to the same lowest order in the LCA and no normalization is artificially imposed.} We find sub-percent deviations that can be attributed to the truncation of the summation over the quantum numbers $\left( U, M_U \right)$ in Eq.~(\ref{eq:LCA_full}) and the introduction of finite grid sizes in $(r,k)$.   Note that in computing matrix elements with $w(r,k)$ one  multiplies it with $r^2 k^2$, see for example Eqs.~(\ref{eq:tkin_wigner}) and (\ref{eq:rms_wigner}).

Inspecting the results of Fig.~\ref{fig:wigner}, one observes that in IPM the $w(r,k)$ extends over the entire radial range and over a well-constrained $k$-range that is almost identical for all three nuclei considered. The SRCs generate a fat momentum tail in $w(r,k)$ that is mainly confined to the interior of the nucleus. Correspondingly, the high-momentum SRC contributions are distributed in a narrower $r$ range than the mean-field contributions.
Indeed, the weight of the fat tails diminishes with increasing $r$, with the largest weight at $r<r_\text{rms}$, and hardly any high-momentum components for $r \gtrsim 2r_\text{rms}$.  This is a reflection of the fact that in LCA the SRCs are chiefly generated from correlation operators acting on IPM nodeless relative $S$-pairs~\cite{Vanhalst:2011es}. Obviously, the wave functions for these $S$-pairs have a finite density at relative $r_{12}=0$ and are very much confined to the nuclear interior.  The panels in the bottom row of Fig.~\ref{fig:wigner} consistently show that the strength of the SRCs contribution to the LCA $w(r,k)$ shifts to smaller $r$ compared to the IPM. Note that the use of the term \emph{SRC} does not imply a momentum cut here. There are significant contributions to the SRCs part from $k<k_F$ as the bottom row in Fig.~\ref{fig:wigner} shows. The normalization of the SRC part (1.8/6 nucleons for $\carbon$) cannot be directly connected with experimental measures for the number of SRC pairs as those exclusively refer to high-momentum SRCs. The maximum of the Wigner distribution (and the rms radius) for the SRCs contribution sits at smaller values of $r$  in comparison to the IPM and LCA ones.  We verified that SRCs make up almost 100\% of the LCA result for $k>2~\text{fm}^{-1}$ at any $r$. As mentioned earlier the absolute SRC contribution diminishes strongly for large $r$ at these higher momenta.  This might raise the question to what extent the SRCs in the interior can be probed in scattering experiments, where the ejected particles are subject to strong final-state interactions (FSI) while traversing the nuclear medium.  A previous study~\cite{Cosyn:2009bi} has shown, however, that reactions probing correlated nucleon  pairs are still sensitive to the nuclear interior, even after correcting for FSI effects.  This is in contrast to single-nucleon knockout reactions which become much more surface dominated after FSI corrections.

For $\carbon$, the two-dimensional Wigner distributions $w(r,k)$ in Fig.~\ref{fig:wigner} has a distinctive negative region for small $r$ and $k$, illustrative of  quantum effects.  For the two calcium isotopes, the area with $w(r,k) <0 $ is also located at small $r$ and $k$. The proton $w(r,k)$ are almost identical in $\calcium$ and $\calciumN$ (note the log scale), whereas for neutrons in $\calciumN$ the emergence of a surface-located neutron skin is clearly visible.  In the forthcoming, it will be shown that the relative impact of SRCs on the proton and neutron kinetic energies differs in an asymmetric nucleus \cite{Frick:2004th,Sargsian:2012sm,Hen:2014nza, Ryckebusch:2018rct}.
Compared to the deuteron results for $w(r,k)$ of Ref.~\cite{Neff:2016ajx}, the fat momentum tails in LCA for finite nuclei extend over a larger momentum range and do not display the oscillations. These differences are likely attributed to the smoothing effect from pair c.o.m. motion and the fact that many pairs are contributing.

 \begin{figure}[htb]
    \centering
    \includegraphics[width=\linewidth]{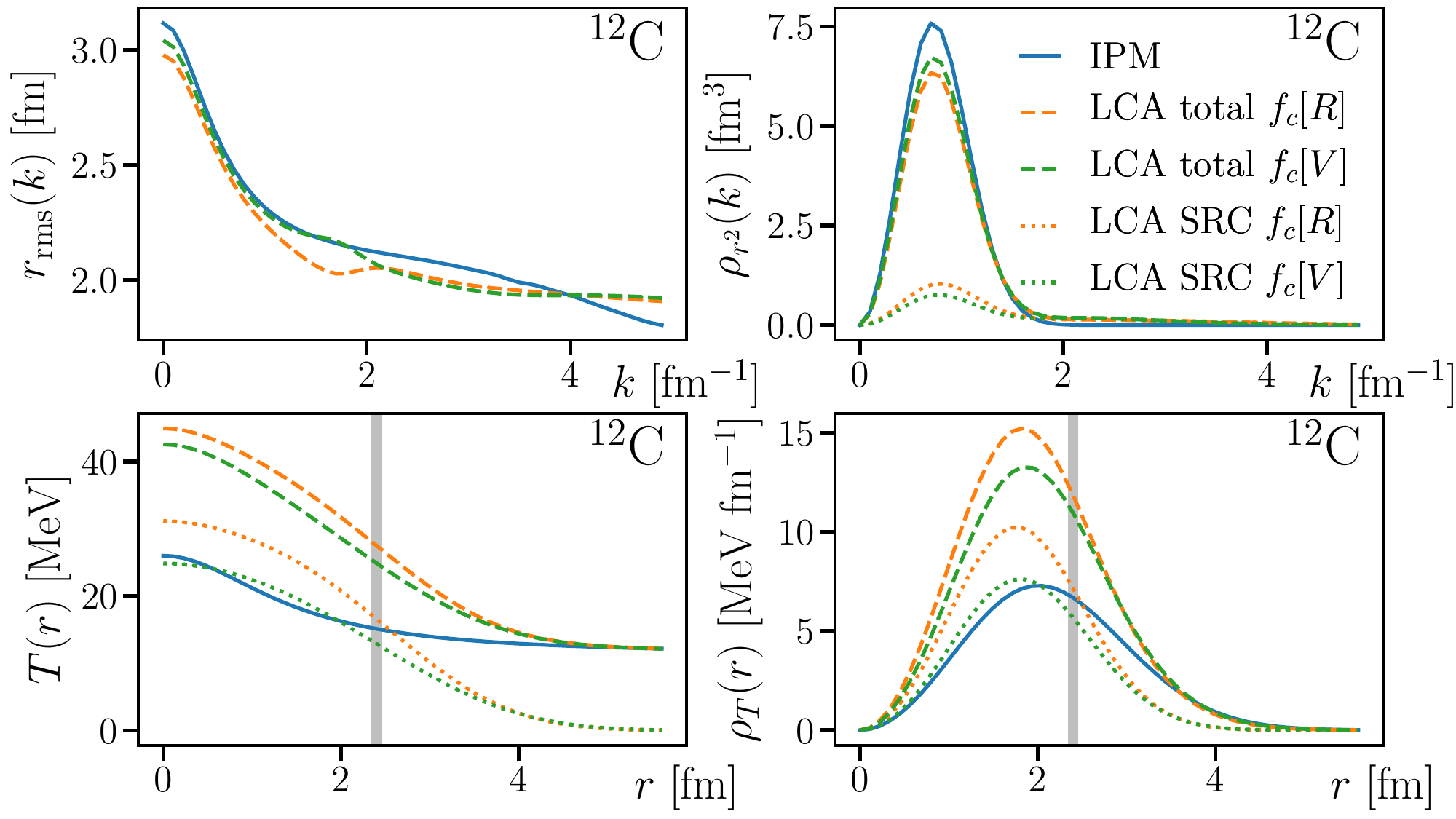}
    \caption{Kinetic energy and rms radius of $^{12}$C.  Top row: momentum dependence of rms radius (all nucleons). Bottom row: radial dependence of kinetic energy per nucleon (all nucleons).  The left column shows the expectation value of the operator at a specific $r$ or $k$ [Eqs.~(\ref{eq:exp_T}), (\ref{eq:exp_r})].  The right column plots the density of the corresponding observable in $r$ or $k$ [Eqs.~(\ref{eq:tkin_wigner}), (\ref{eq:rms_wigner})].  All curves use the $\hbar\omega[d]$ HO inputs.  ``LCA SRC'' curves show the contribution from SRCs to the ``LCA total'' result.  The difference between the two is not equal to the IPM curve because both the IPM and LCA total result use distributions normalized to the total number of nucleons.  
    The grey vertical bands in the lower row cover the range of point rms radii for the depicted model inputs, see Table~\ref{tab:rmskin}. }    \label{fig:C12_obs}
\end{figure}

\begin{figure}[htb]
    \centering
    \includegraphics[width=\linewidth]{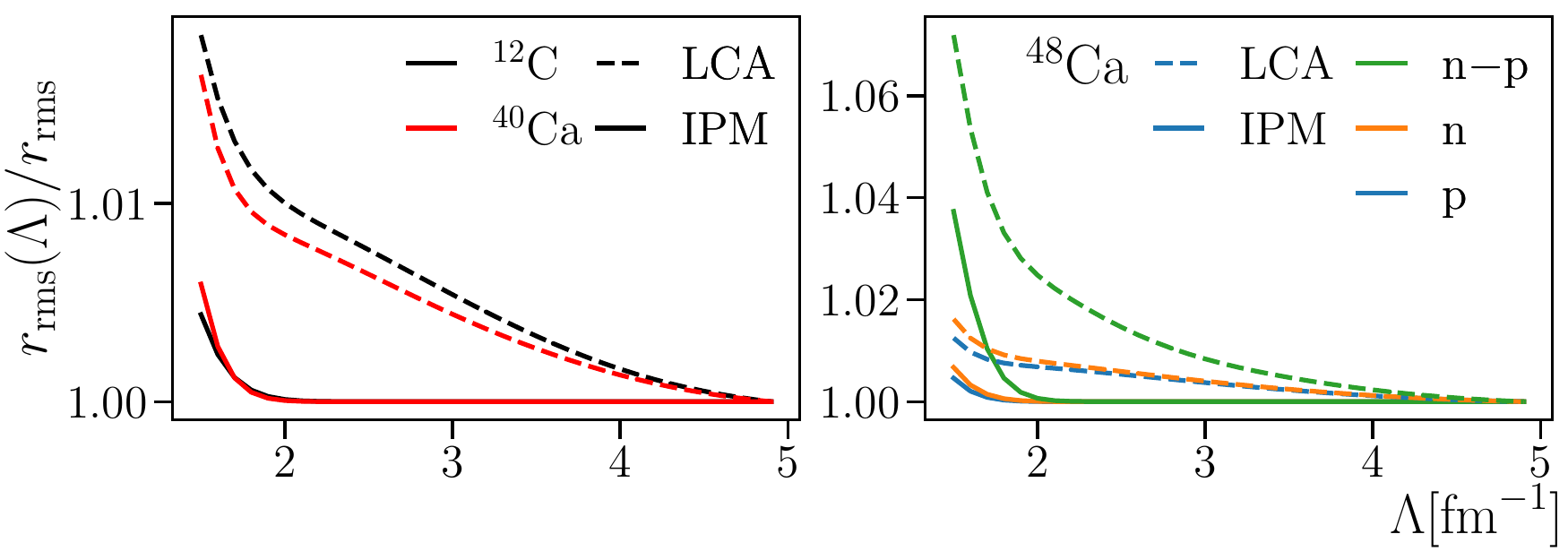}
    \caption{Momentum cutoff dependence of rms radii.  We show the ratio of point rms radii calculated with a high-momentum cutoff $\Lambda$ to their $\Lambda = +\infty$ values.  $\carbon$ and $\calcium$ are shown in the left panel; $\calciumN$ proton, neutron and neutron skin (n$-$p) values in the right panel. IPM curves use the $\hbar\omega[d]$ input, LCA curves use $\hbar\omega[f]/f_c[R]$. Other IPM/LCA inputs produce similar curves. }
    \label{fig:rms_cutoff}
\end{figure}

We now discuss the numerical results for the Wigner-based functions that are introduced to investigate the spatial dependence of the kinetic energy ($T(r)$ and $\rho_T(r)$ of Eqs.~(\ref{eq:exp_T}) and (\ref{eq:tkin_wigner})) and the momentum dependence of the nuclearradii ($r_{\text{rms}}(k)$ and $\rho_{r^2}(k)$ of Eqs.~(\ref{eq:exp_r}) and (\ref{eq:rms_wigner})).   In what follows, we systematically compare IPM with LCA results and we  stress that both are obtained for identical normalizations of the $w(r,k)$.
The $\carbon$ results for the $\hbar\omega[d]$ model variants of Table~\ref{tab:fits} are shown in Figure~\ref{fig:C12_obs}. Similar results are obtained with the $\hbar\omega[f]$ HO frequency. The SRCs substantially increase the $T(r \lesssim 4 ~\text{fm})$ and $\rho _{T}(r \lesssim 4 ~\text{fm})$, whereby the effect gradually decreases with growing radial distance $r$.   For $r>4$ fm, the IPM and LCA $T(r)$ and $\rho _{T}(r)$ coincide, showing again the minor role played by SRCs beyond the high-density regions in the nuclear interior.  The uncertainty on $T(r)$ connected with the choice of the central correlation function is largest for $r=0$ and is of the order of 20\% on the SRC contribution and 10\% on the LCA result.  For $r<r_\text{rms}$ the SRCs account for roughly 60-70\% of the kinetic energy.  The $\rho_T(r)$ demonstrates that the kinetic energy receives significant contributions from $r>r_\text{rms}$.  For the rms radius, the overall impact of the SRCs is far more modest than what is observed for the kinetic energies.  The LCA predictions for $\rho_{r^2}(k)$  are below the IPM one except for $k>4~\text{fm}^{-1}$.  Overall this implies smaller radii when including SRCs using the same HO frequency, as was shown in Fig~\ref{fig:rms_fit}. From the $\rho_{r^2}(k)$ one infers that both the IPM and LCA  contributions to $r_\text{rms}$ from $k>2~\text{fm}^{-1}$ are very small despite the fact that the LCA $\rho_{r^2}(k)$ is much larger than the IPM one for high momenta.  We look at these high-momentum contributions in more detail in Fig.~\ref{fig:rms_cutoff}.

\begin{figure}[htb]
    \centering
    \includegraphics[width=\linewidth]{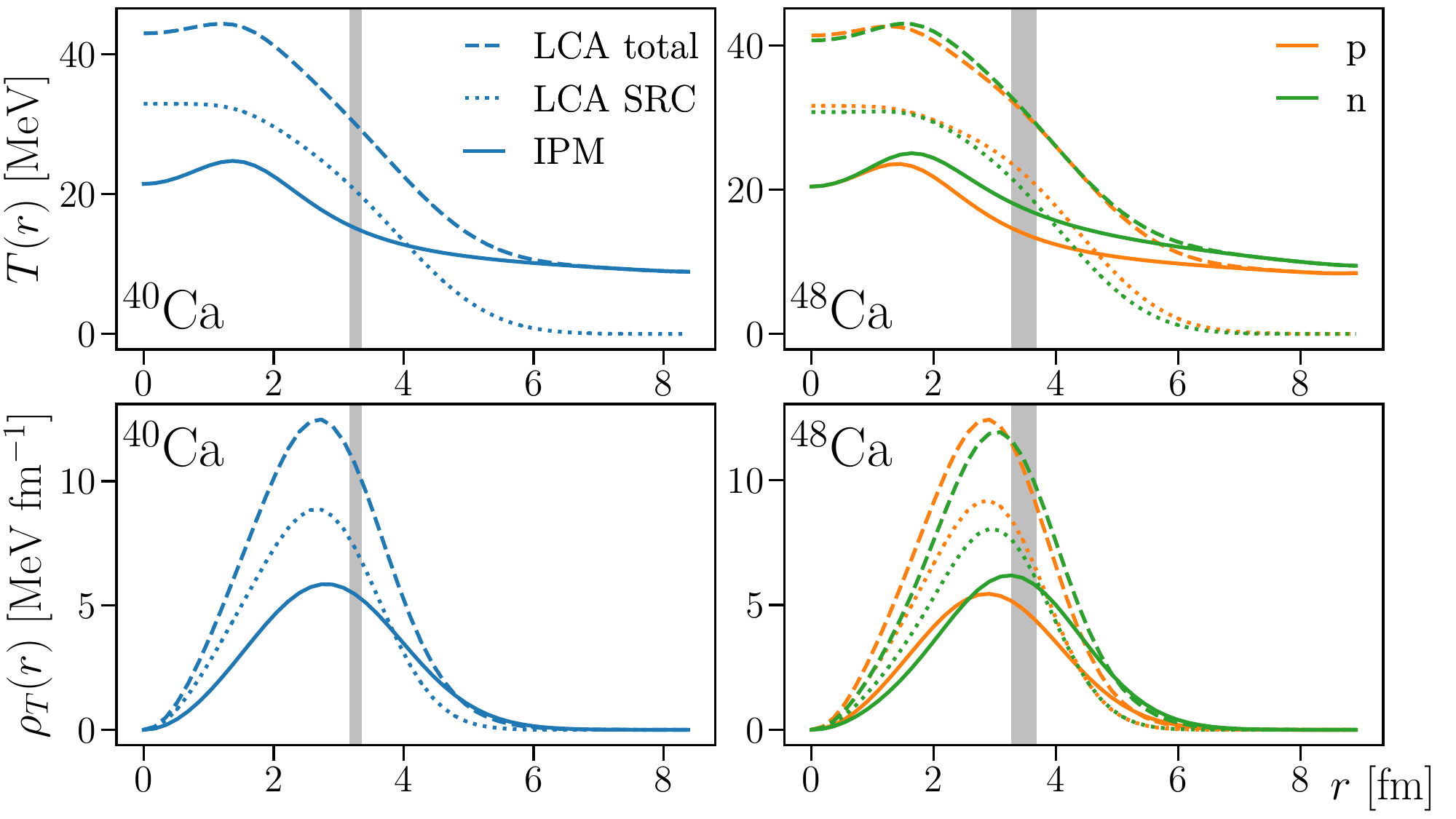}
    \caption{$\calcium$ and $\calciumN$ radial dependence of kinetic energy expectation value $T(r)$ (top row) and density $\rho_T(r)$ (bottom row). All LCA results use the $\hbar\omega[d]/f_c[R]$ input, IPM uses $\hbar\omega[d]$. Proton and neutron results are identical in $\calcium$. Normalization of ``LCA SRC'' as in Fig.~\ref{fig:C12_obs}.   Legends (linestyle/color) apply to all panels.  The vertical bands cover the range of point rms radii for the depicted models, see Table~\ref{tab:rmskin}.}
    \label{fig:Ca_obs_T}
\end{figure}

\begin{figure}[htb]
    \centering
    \includegraphics[width=\linewidth]{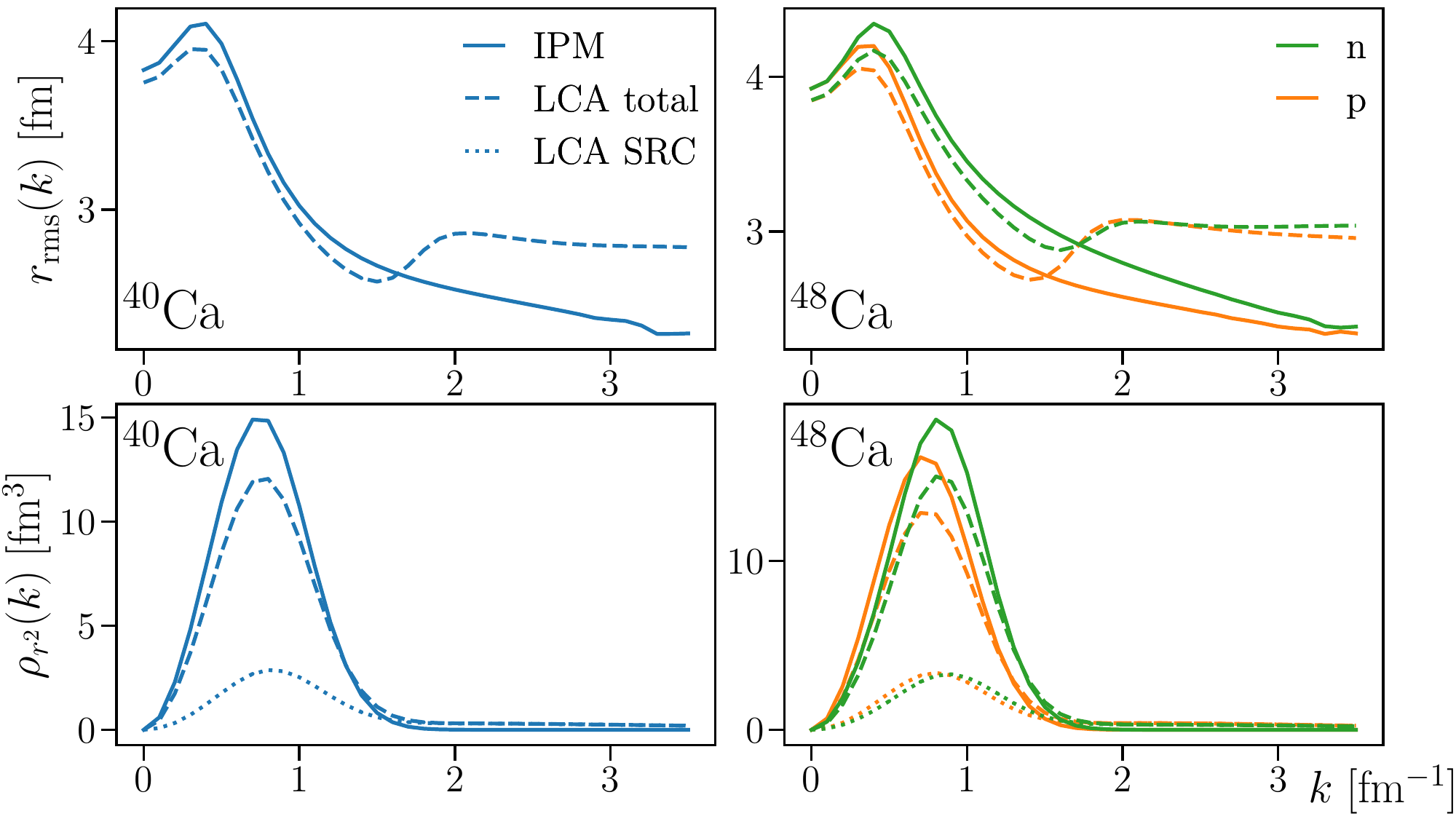}
    \caption{As in Fig.~\ref{fig:Ca_obs_T} but now for the momentum dependence of the $r_\text{rms}(k)$ expectation value (top row) and $\rho_{r^2}(k)$ density (bottom row).
}
    \label{fig:Ca_obs_r}
\end{figure}

In Figs.~\ref{fig:Ca_obs_T} and \ref{fig:Ca_obs_r}, IPM and LCA results for  the combinations $\left[T(r), \rho_T(r) \right]$ and $ \left[ r_{\text{rms}}(k), \rho_{r^2}(k) \right]$ for $\calcium$ and $\calciumN$ are shown. Our major focus here is how SRCs impact the proton and neutron features and the role of the neutron excess in $\calciumN$. To this end we discuss the numerical results of the IPM model with the $\hbar\omega[d]$ HO frequency and compare them with results of the LCA model with the $\hbar\omega[d]/f_c[R]$ input.  In general, similar trends as for the $\carbon$ results of Fig.~\ref{fig:C12_obs} are observed. Comparing the radial dependence of the kinetic energy $T(r)$ for the protons and neutrons in $\calciumN$ in Fig.~\ref{fig:Ca_obs_T}, we can make the following observations.  In the IPM neutrons possess more kinetic energy than protons for all $r$.  For the SRC contribution to the LCA result, the situation is reversed. As far as the SRCs contributions are concerned, protons are more kinetically energetic for all $r$.   The weighted sum of the IPM and SRCs contributions results in the radial dependence of the LCA $T(r)$.  In the deep interior $r \lesssim 1.8$~fm of the nucleus, where the SRC effects are in full swing, protons have more kinetic energy, while for $r \gtrsim 1.8$ fm neutrons have more.  The difference is very small, and   in the deep surface region $r>6$ fm there is no effect from the SRC.  The $\calciumN$ $\rho_T(r)$ shows that counted per nucleon the IPM protons in the nuclear interior contribute more to the bulk kinetic energy than the neutrons, while the situation is reversed in the exterior. In the LCA, the first effect is enhanced while the second is reduced.  The combination of these effects results in more kinetically energetic protons than neutrons after including SRCs (see Table~\ref{tab:rmskin}).

In line with the carbon results of Fig.~\ref{fig:C12_obs}, the radius $r_\text{rms}(k)$ of the calcium isotopes in Fig.~\ref{fig:Ca_obs_r} reaches a plateau at $k \gtrsim 2~\text{fm}^{-1}$. The $\calciumN$ neutron radii are larger than the proton ones for momenta below the Fermi one, but in the momentum region dominated by the tensor correlations ($1.5~\text{fm}^{-1} \lesssim k \lesssim 2.5~\text{fm}^{-1}$) the LCA proton and neutron radii almost coincide.   As with $\carbon$,  the $\rho_{r^2}(k)$ indicate that high-momentum nucleons only contribute marginally to the rms radius, even in the LCA.

The resulting kinetic energies $T$ and rms radii $r_\text{rms}$ are summarized in Table~\ref{tab:rmskin} for the different model variants.  The mentioned values are directly obtained through phase-space integration of the $w(r,k)$ (see Eqs.~(\ref{eq:tkin_wigner}) and (\ref{eq:rms_wigner})). 
The $^{40}$Ca results can be compared to those of recent \textit{ab initio} calculations (Tables V and VI of~\cite{Lonardoni:2017egu}). The calculations with the AV18 two-body force lead to $T=32.29~\text{MeV}$ and $r_{\text{rms}}=3.41~\text{fm}$. In LCA we find values of $T$ and $r_{\text{rms}}$ that are slightly smaller but the deviation is at most 10\%. We stress that in LCA we do not have long-range correlations. In that light it is important to note that the results of \cite{Lonardoni:2017egu} indicate that three-body forces decrease the $T$ by about 1~MeV and increase the $r_{\text{rms}}$ by about 0.1~fm.

For $\calciumN$, one observes that all LCA model variants systematically predict that the proton kinetic energy is about 1~MeV larger than the neutron one, a phenomenon known as ``kinetic energy inversion''~\cite{Sargsian:2012sm}.  For the rms radii, the LCA result shows a slight reduction of the different radii for the $\hbar\omega[d]$ inputs, while SRC connected changes in the size of the $\calciumN$\ neutron skin are of the order 5-10\%.  
Note that the kinetic energy inversion for protons and neutrons in $\calciumN$ cannot be directly inferred from the shapes of the $w(r,k)$ of Fig.~\ref{fig:wigner}.  Indeed, the difference between $\calciumN$ proton and neutron values for $k_\text{rms}$ (which is related to $\langle T \rangle$ as $\langle T \rangle = k_\text{rms}^2/2m$) is hardly visible in Fig.~\ref{fig:wigner}.

Overall, the impact of the SRCs relative to the IPM result for the radii and kinetic energies is rather insensitive to the input choices of the calculations. For the radii the largest uncertainties stem from the choice of the HO parameter. As illustrated by the $\hbar \omega [d]$ results that use a fixed HO frequency, the impact of SRCs  is a reduction of the order 3-6\%. The SRCs systematically reduce the neutron skin. For the kinetic energies, the impact of the SRCs is large and the uncertainty stemming from the choice for the central correlation functions on the LCA result is of the order of 2-3~MeV. The choice for the oscillator parameter has a somewhat smaller (1 to 2~MeV) impact on the LCA kinetic energies.

\begin{table*}[htb]
    \centering
    \caption{Computed kinetic energies per nucleon (in MeV) and point-nucleon rms radii (in fm). The comparison between the computed and measured charge radii for the different model variants is displayed in Fig.~\ref{fig:rms_fit}.}
    \label{tab:rmskin}
\resizebox{0.6\textwidth}{!}{    \begin{tabular}{l|cc|cc|cccccc}
\toprule
          & \multicolumn{2}{c|}{$\carbon$} & \multicolumn{2}{c|}{$\calcium$} &  \multicolumn{6}{c}{$\calciumN$}   \\
         Model &$T_{p,n}$ & $r_{p,n}$ & $T_{p,n}$ & $r_{p,n}$ & 
         $T_p$ & $T_n$ & $T_n$-$T_p$ &
         $r_p$ & $r_n$ & $r_n$-$r_p$\\
          \midrule
        IPM $\hbar \omega[d]$ &   16.1 & 2.46 & 16.5 & 3.36 & 
        15.7 & 18.0 & 2.2 &
         3.44 & 3.68 & 0.237\\

        LCA $\hbar \omega[d]$/$f_c[R]$ & 29.7 & 2.34 & 31.9 &3.17 & 
         32.5 & 31.6 & -1.0 &
        3.27 & 3.48 & 0.216\\
        LCA $\hbar \omega[d]$/$f_c[V]$  & 26.8 & 2.40 & 28.9 & 3.22 & 
        29.5 & 28.7 & -0.8 &
        3.31& 3.53 & 0.221\\
        IPM $\hbar \omega[f]$ & 18.3 & 2.30 & 17.2 & 3.29 &
        16.2 & 18.5 & 2.3 &
        3.39 & 3.63 & 0.234 \\
        
        LCA $\hbar \omega[f]$/$f_c[R]$ & 30.4 & 2.31 & 30.1 & 3.28 &
        30.5 & 29.5 &-0.9 &
        3.39 & 3.62 & 0.226\\
        LCA $\hbar \omega[f]$/$f_c[V]$ & 28.7 & 2.32 & 27.8 & 3.28 &
        28.2 & 27.5 & -0.8 &
        3.39 & 3.62 & 0.227 \\
        \bottomrule
    \end{tabular}}
\end{table*}

So far, no momentum selection was imposed when investigating the influence of SRCs on bulk nuclear properties.  In order to keep the dimensionality of the numerical \emph{ab-initio} calculations in check one often relies on softened nucleon-nucleon interactions that involve a momentum cut-off. In light of this, we wish to quantify the influence of SRCs on the nuclear radii as a function of a momentum cut-off. 
To this end, we introduce a momentum cutoff $\Lambda$ in the momentum integrals of Eq.~(\ref{eq:rms_wigner}), both in the numerator and denominator.  We show the relative change in the rms radii as a function of $\Lambda$ in Fig.~\ref{fig:rms_cutoff}.  The IPM (LCA) radii converge at $\Lambda \approx 2~\text{fm}^{-1}$ ($\Lambda \approx 4-5~\text{fm}^{-1}$). The momentum range $2 \lesssim k \lesssim 4~\text{fm}^{-1}$ shrinks the LCA rms radii by about 1\%. This might seem to be at odds with the obvious fact that at large momentum the LCA $r_{\text{rms}} (k)$ of Fig.~\ref{fig:Ca_obs_r} overshoots the IPM one. These $r_\text{rms}(k>2~\text{fm}^{-1})$ values, however, are smaller than those for $r_\text{rms}(k<2~\text{fm}^{-1})$. This means that excluding the $k>2~\text{fm}^{-1}$ (see $\rho_{r^2}(k)$ in Figs.~\ref{fig:C12_obs} and \ref{fig:Ca_obs_r}), increases the LCA rms radii.  For $\calciumN$ the per nucleon influence of SRCs is different for protons than for neutrons, resulting in a slight decrease of the neutron skin on the order of 3\% when including high-momentum SRCs with $k>2~\text{fm}^{-1}$, independent of the chosen LCA model.  This agrees with the naive picture of the dominant tensor correlations pulling the high-momentum protons close to the neutrons, see $r_\text{rms}(k)$ in Fig.~\ref{fig:Ca_obs_r}.

\section{Conclusion and Outlook}
\label{sec:conclusion}
%

Most often nuclear short-range correlations in finite nuclei are addressed from the perspective of distributions in momentum space and the operational definition of SRCs is one whereby one refers to nucleons with a momentum well above the Fermi momentum ($k \gtrsim k_F = 1.2~\text{fm}^{-1}$). This implies that the SRC terminology is exclusively used for nucleons that have a momentum larger than $k_F$.  In this work, we have added the spatial perspective by presenting calculations of nuclear Wigner distributions $w(r,k)$ that include the effect of SRCs.  The calculations are conducted within  the lowest-order correlation operator approximation (LCA), a framework that has shown its interpretative potential in dealing with observations.

Common features emerged for the $w(r,k)$ of the three nuclei $\carbon$, $\calcium$ and $\calciumN$ considered in this work.  The SRCs impact the $w(r,k)$ over the full momentum range considered  (from 0 up to the nucleon mass) and a radial range that is confined to the deep nuclear interior. The SRC contribution to the Wigner distribution has a $k_{\text{rms}}$ of about 1.6-1.7~fm$^{-1}$, whereas the IPM $w(r,k)$ has $k_{\text{rms}} \approx 0.8$~fm$^{-1}$. The resulting LCA $w(r,k)$ that has both mean-field and SRC contributions has  $k_{\text{rms}} \approx 1.2$~fm$^{-1}$.  As the SRCs are confined to the deep interior of the nucleus, the LCA $w(r,k)$ has an $r_{\text{rms}}$ that is approximately 0.2~fm smaller than the IPM one when using the same HO frequency.  

From the $w(r,k)$ we computed a prototypical bulk nuclear momentum-space and coordinate-space feature, namely the non-relativistic kinetic energy and the point-nucleon radius. The SRCs almost double the proton and neutron kinetic energies. Furthermore, the proton-neutron dominance of the SRCs gives rise to peculiar effects in asymmetric nuclei with a neutron abundance whereby the per-nucleon kinetic energy is larger for protons than for neutrons. The impact of SRCs on proton and neutron radii is at the percent level. In asymmetric nuclei the proton and neutron radii are impacted differently. In the asymmetric nucleus $\calciumN$ studied here, the SRCs reduce the neutron skin by an amount that is of the order of 5-10\%, with high-momentum SRC components accounting for 3\%.     

The Wigner distributions discussed here could provide valuable input in semi-classical transport calculations for nuclear reaction cross sections \cite{Buss:2011mx,Golan:2012wx,Andreopoulos:2015wxa} that wish to account for the effects of SRCs.  It would be interesting to see whether a model with a realistic description of both long-range and short-range correlations leads to sizable changes in the nuclear radii.

\vspace{0.5em}
The computational resources (Stevin Supercomputer Infrastructure) and services used in this work were provided by the VSC (Flemish Supercomputer Center), funded by Ghent University, FWO and the Flemish Government – department EWI.  We thank M. Sargsian for comments on an earlier draft.

\appendix
\section{Wigner distribution in the LCA}
\label{app:lca_wigner}

 For details of the formalism underlying the LCA, we refer to Refs.~\cite{Vanhalst:2011es,Vanhalst:2012ur, Vanhalst:2014cqa, Ryckebusch:2018rct, Ryckebusch:2019oya}.
Here, we outline the major derivation steps of the nuclear Wigner distribution in LCA.
In the LCA, normalized and antisymmetrized two-particle HO states with quantum numbers $\alpha_i \equiv n_il_ij_im_{j_i}t_i (\bm r_i)$ are expanded in coupled relative and center-of-mass HO states using Moshinsky brackets  
\begin{equation}\label{eq:HOtwobodystates}
    \bigl|\alpha_1\alpha_2\bigr\rangle = C^\mathcal{A}_{\alpha_1\alpha_2}\Bigl| \mathcal{A}\equiv n(lS)jm_j(\bm r_{12}),NLM_L(\bm R_{12}),TM_T\Bigr \rangle\,.
\end{equation}
See Ref.~\cite{Vanhalst:2011es} for the full expressions of $C^\mathcal{A}_{\alpha_1\alpha_2}$.
Through the action of universal SRC operators, one-body operators become effective two-body operators in the LCA.  The two-dimensional Wigner distribution of Eq.~(\ref{eq:wignerinrandk}) can be computed with the two-body operator 
\begin{multline} \label{eq:w_LCA_operator}
    \widehat{w}_{t_1 t_2}(r_1,k_1) = 
    \frac{1}{(2\pi)^3} \sum_{s_1,s_2}\iint 
    d\Omega_{\bm r _1}d\Omega_{\bm k _1}
    \iint d \bm q\, d\bm k_2 \; \\\times
    e^{-i\bm q\cdot \bm r_1}\; \biggl| \bm k_1 + \frac{\bm q}{2},s_1t_1;\bm k_2, s_2t_2 \biggr\rangle \biggl\langle \bm k_1 + \frac{\bm q}{2},s_1t_1;\bm k_2, s_2t_2 \biggr| \,.
\end{multline}
Note that the separate nucleon-pair contributions $t_1 t_2 \in \{ pp,nn,np,pn \}$ to the matrix element are calculated.  The $(pp,pn)$ and $(nn,np)$ terms are combined to obtain the proton and neutron Wigner distributions respectively. The LCA matrix elements between coupled states adopt the form
\begin{multline}\label{eq:LCA_me}
    w^{\mathcal{A}\mathcal{A'},\text{LCA}}_{t_1t_2}(r_1,k_1) = \Bigl\langle \mathcal{A}\,\Bigr| \left[f_p(r_{12})\widehat{O}^p(1,2)\right]^\dagger  \\\times
   \widehat{w}_{t_1 t_2}(r_1,k_1)
    \left[f_q(r_{12})\widehat{O}^q(1,2)\right] \Bigl|\,\mathcal{A}'\Bigr\rangle\,,
\end{multline}
where $f_{p,q}(r_{12}) \widehat{O}^{p,q}(1,2)$ are one of the three correlation operators (central, tensor, spin-isospin) introduced in LCA.  The strength functions $f_i(r_{12})$ depend on the internucleon distance. Note that an IPM calculation of the two-dimensional Wigner distribution of Eq.~(\ref{eq:LCA_me}) corresponds with replacing the correlation operator with the unity one.  The effect of the SRCs on the overall normalization is accounted for by dividing LCA matrix elements by a norm calculated to the same order using Eq.~(\ref{eq:LCA_me}) with $\widehat{w}(r_1,k_1) \rightarrow 1$.

For the three included SRC operators, the action of the correlation operators on the coupled states can be schematically written as
\begin{multline}\label{eq:decomposition}
    \widehat{O}^i(1,2) \Bigl| \mathcal{A}\equiv n(lS)jm_j(\bm r_{12}),NLM_L(\bm R_{12}),TM_T\Bigr \rangle \\
      =\sum_{l'=|j-1|}^{j+1} O^i(S,T,j,l,l')\\\times\Bigl|nl(l'S)jm_j(\bm r_{12}),NLM_L(\bm R_{12}),TM_T \Bigr \rangle\,.
\end{multline}
The included correlation operators can only change the quantum number $l$ (to $l'$) in the angular part of the relative HO coordinate, while in the radial part of the HO wave function it remains unchanged. No other quantum numbers change. The coefficients $O^i$ in the decomposition of Eq.~(\ref{eq:decomposition}) depend on $(S,T,j,l,l')$, and only the tensor correlation operator generates coefficients with $l\neq l'$.
The final result is obtained 
using the following steps:
\begin{enumerate}\setlength\itemsep{-0.3em}
    \item Inserting two complete sets of coordinate space wave functions $(\bm r_1,\bm r_2)$ in Eq.~(\ref{eq:LCA_me}) and rewriting everything in relative $(\bm r_{12})$ and c.o.m. $(\bm R_{12})$ coordinates.
    \item Inverse Fourier transforming the radial HO wave functions in the c.o.m. coordinates.
    \item Expanding the plane waves in terms of Bessel functions and spherical harmonics.
    \item Making use of spherical harmonic identities.
\end{enumerate}
One obtains
\begin{multline} 
     w^{\mathcal{A}\mathcal{A'},\text{LCA}}_{t_1t_2}(r_1,k_1) = \sum_{l_p = 
|j-1|}^{j+1} \sum_{l'_q = |j' - 1|}^{j'+1} \text{O}^{p\dagger}(S,T,j,l,l_p) \\
\times\text{O}^{q}(S,T',j',l',l_q')
\langle \tfrac{1}{2}t_1\tfrac{1}{2}t_2|T M_T\rangle 
\langle \tfrac{1}{2}t_1\tfrac{1}{2}t_2|T' M'_T\rangle\\
\times \sum_{m_S}\langle l_p m_{l_p} S m_S|j m_j\rangle 
\langle l'_q m_{l'_q} S m_S|j' m'_j\rangle \\
\times \frac{2^5}{\pi^2}\sum_{u m_u}\sum_{K m_K} \sum_{K' m'_K} i^{L-L'-K+K'}
 \threej{ l_p,u,K,m_{l_p},m_u,m_K} \\
 \times
 \threej{ l_p,u,K,0,0,0}  
 \threej{ l_q',u,K',m_{l_q'},m_u,m_{K'}} 
 \threej{ l_q',u,K',0,0,0} \\
  \times\sum_{U m_U} \hat{u}^{2} \hat{U}^{2} \hat{l}_{p} \hat{l}_{q}' \hat{k}^{2} \hat{k}'^{2} 
\hat{L} \hat{L}' 
 \threej{ L,K',U,M_L,m'_{K},m_U} \\
 \times
 \threej{ L,K',U,0,0,0} 
 \threej{ L',K,U,M_{L'},m_K,m_U} 
 \threej{L',K,U,0,0,0} \\
 \times
\xi_{p,nlN'L'}^{UuK}(r_1,k_1)
\xi_{q,n'l'NL}^{UuK'}(r_1,k_1)\,,
\label{eq:LCA_full}
\end{multline}
where we used the notation $\hat{j}\equiv \sqrt{2j+1}$. In Eq.~(\ref{eq:LCA_full}), the correlation strength functions $f_i$ enter through
\begin{align}
    &\xi_{i,nlN'L'}^{UuK}(r_1,k_1) = (-1)^{N'}\int dP P^2 \Pi_{N'L'}(P)\nonumber\\
    &\qquad\qquad\qquad\qquad\times j_U(\sqrt{2}Pr_1)\chi_{i,nl}^{uK}(k_1,P)\,,\\
    &\chi_{i,nl}^{uK}(k_1,P) = \int dr r^2 f_i(r)R_{nl}(r)j_u(\sqrt{2}k_1r)j_K(Pr)\,.
\end{align}
The $R_{nl}, \Pi_{NL}$ are defined through the radial parts of the normalized HO wave function $\psi_{nlm_l}(\bm r)$ and its Fourier transform $\phi_{nlm_l}(\bm k)$
\begin{align}
    \psi_{nlm_l}(\bm r) &= R_{nl}(r)Y_{lm_l}(\Omega_r)\,,\nonumber\\
    \phi_{nlm_l}(\bm k) &= \frac{1}{(2\pi)^\frac{3}{2}}\int d^3 r e^{-i\bm k\cdot \bm r}\psi_{nlm_l}(\bm r) \nonumber\\ &= i^{-l}(-1)^n\Pi_{nl}(k)Y_{lm_l}(\Omega_p)\,.
\end{align}
In the numerical evaluation of Eq.~(\ref{eq:LCA_full}), the summation over $U$ is truncated after $ U > 10$, while for the other variables all combinations with non-vanishing $3j$-symbols are retained.  This truncation in the variable $U$ results in normalization errors on the sub-percent level for heavy nuclei.

\bibliographystyle{elsarticle-num}
\bibliography{biblioradius2021.bib}

\end{document}